\documentclass[]{article}

\usepackage{amsmath}
\usepackage{amssymb}

\usepackage[utf8]{inputenc}
\usepackage[T1]{fontenc}

\newtheorem{theorem}{Theorem}[section]

\newtheorem{corollary}{Corollary}[section]
\newtheorem{prop}{Proposition}[section]
\newtheorem{definition}{Definition}[section]

\numberwithin{equation}{section}

\begin{document}
	
\begin{center}
{\bf Pauli Matrices: \\
	A Triple of Accardi Complementary 
	Observables}
\vskip 0.1cm
Stephen Bruce Sontz
\\
Centro de Investigaci\'on en Mat\'ematicas 
\\
(CIMAT)
\\
Guanajato, Mexico
\end{center}

\begin{abstract}
	\noindent 
	The definition due to Accardi of a pair of  
	complementary 
	observables is adapted to the context of the Lie
	algebra $ su(2) $. 
	We show that the pair of Pauli matrices $ A,B $
	associated
	to the unit 
	directions $ \alpha $ and $ \beta $ in 
	$ \mathbb{R}^{3} $ are Accardi complementary 
	if and only if $ \alpha $ and $ \beta $ 
	are orthogonal if and only if $ A $ and $ B $ 
	are orthogonal. 
	In particular, any pair of the 
	standard triple of Pauli 
	matrices is complementary.  
\end{abstract}

\section{Introduction}

The idea of complementarity 
has hung around quantum theory from its
earliest days. 
However, the exact encoding of that 
idea into a rigorous, mathematical 
definition has been elusive. 
A very interesting definition comes from
Accardi's paper \cite{accardi}. 
We extend this definition of 
complementarity from pairs 
of observables to any set, 
finite or infinite, of observables. 
See \cite{accardi-lu} and \cite{cas-vara} 
for more on this topic.

\section{Preliminaries}
\label{Prelim-section}

In this section we review some 
standard material in order to establish notation 
and context. 
We use the standard notation for the Pauli matrices:
$$
\sigma_1 = \left( \begin{array}{cc} 0 & 1 \\ 1 & 0 \end{array} \right),
\qquad
\sigma_2 = \left( \begin{array}{cc} 0 & - i \\ i & 0 \end{array} \right),
\qquad
\sigma_3 = \left( \begin{array}{cc} 1 & 0 \\ 0 & -1 \end{array} \right).
$$
For any vector $ \alpha = 
(\alpha_{1}, \alpha_{2}, \alpha_{3})
\in \mathbb{R}^{3} $
we let
$$
   \alpha \cdot \sigma := \alpha_{1} \sigma_{1} +
   \alpha_{2} \sigma_{2} + \alpha_{3} \sigma_{3} =
   \left( 
       \begin{array}{cc} 
           \alpha_{3} & \alpha_{1} - i \alpha_{2}  \
           \\ 
           \alpha_{1} + i \alpha_{2}  & -\alpha_{3} 
       \end{array} 
    \right).
$$
We denote the unit sphere in $ \mathbb{R}^{3} $
as
$$
S^{2} := \{ \alpha = 
(\alpha_{1}, \alpha_{2}, \alpha_{3})
\in \mathbb{R}^{3} ~\big|~ || \alpha ||^{2} :=
\alpha_{1}^{2} + \alpha_{2}^{2} + \alpha_{3}^{2} = 1
\}. 
$$
For a $ 2 \times 2 $ matrix $ A $ we 
will use the 
normalized trace, $ \mathrm{tr} \, A 
: = (a + d)/2 $, 
where 
$$
A = \left( \begin{array}{l} a ~~ b \\ c ~~ d \end{array} \right)
\qquad \mathrm{for~} a,b,c,d \in \mathbb{C}.
$$
Since 
$ \sigma_{1}^{2} = \sigma_{2}^{2} =\sigma_{3}^{2} = I $, 
where $ I $ is the $ 2 \times 2 $ identity matrix, 
we have that 
$ \mathrm{tr} \,\sigma_{1}^{2} = 
\mathrm{tr} \,\sigma_{2}^{2} =
\mathrm{tr} \,\sigma_{3}^{2} = 1 $. 
Also, 
$ \mathrm{tr} \,\sigma_{1} = 
\mathrm{tr} \,\sigma_{2} =
\mathrm{tr} \,\sigma_{3} = 0 $.
We denote the set of hermitian (i.e., self-adjoint), 
traceless $ 2 \times 2 $ matrices 
with complex entries by 
$$ 
su (2) := \{ A~|~ A = A^{*}, \, 
\mathrm{tr} \, A =0 \}.   
$$
Here $ M^{*} $ denotes the adjoint (complex conjugate,
transposed) of $ M $, where $ M $ is any matrix, 
even a rectangular one. 
Then $ su (2) $ 
is the Lie algebra of the Lie group $ SU(2) $.
It is a vector space over the reals $ \mathbb{R} $,
and the map 
$\Sigma : \mathbb{R}^{3} \to su(2)  $
given for each $ \alpha \in \mathbb{R}^{3} $
by $ \Sigma (\alpha) :=  \alpha \cdot \sigma $
is a linear, onto isomorphism of real vector spaces. 
In particular, for 
all $ \alpha \in \mathbb{R}^{3} $
we have 
$ (\alpha \cdot \sigma)^{*} = \alpha \cdot \sigma $ 
and 
\begin{equation}
\label{traceless}
\mathrm{tr} \, (\alpha \cdot \sigma) = 0.
\end{equation} 

Moreover, we give $ \mathbb{R}^{3} $ the standard 
inner product, denoted 
$ \langle \cdot , \cdot \rangle  $, 
and we give $ su(2) $
the inner product, 
also with the same notation, by restricting to $ su(2) $ 
the normalized Hilbert-Schmidt inner product: 
$$
   \langle A, B \rangle := \mathrm{tr} \, ( A^{*} B ) 
   \quad \mathrm{for~all~}2 \times 2
   \mathrm{~matrices~}
    A,B. 
$$
Then $ \Sigma $ is a unitary isomorphism 
of $ \mathbb{R}^{3} $ onto $ su(2) $. 
Explicitly, 
this says that
\begin{equation}
\label{unitary}
\langle
\Sigma(\alpha), \Sigma(\beta)
\rangle = 
\langle 
\alpha \cdot \sigma, \beta \cdot \sigma  
\rangle = 
\mathrm{tr} 
\big( (\alpha \cdot \sigma) (\beta \cdot \sigma) \big) 
= \langle \alpha , \beta \rangle
\quad \mathrm{for~all~} 
\alpha, \beta \in \mathrm{R}^{3}. 
\end{equation}
Also, 
$\{ \sigma_{1} , \sigma_{2} , \sigma_{3} \} $
is an orthonormal basis of $ su(2) $. 
The matrices in $ su(2) $, which are all self-adjoint, 
act on the
Hilbert space $ \mathbb{C}^{2} $ with its standard 
inner product and, as such, represent quantum physical 
observables. 

Using this notation and standard properties of the
Pauli matrices, we have 
$
(\alpha \cdot \sigma)^{2} = || \alpha ||^{2} \, I.
$
From this one immediately has that the spectrum 
of $ \alpha \cdot \sigma $ is
$
   \mathrm{Spec} (\alpha \cdot \sigma) = 
   \{\, -|| \alpha ||^{2}, || \alpha ||^{2} \, \}. 
$
This motivates calling $ \alpha \cdot \sigma $ 
for $ \alpha \in S^{2} $
the {\em Pauli matrix in the direction $ \alpha $}. 
Also, in quantum 
physics $ (\hbar/2) (\alpha \cdot \sigma) $ is
called the 
{\em spin matrix in  the direction $ \alpha $}, 
where $ \hbar > 0 $ is the normalized Planck constant. 
The results of this paper are given in terms of 
the Pauli matrices $ \alpha \cdot \sigma $ 
for $ \alpha \in S^{2} $. 
However, they are easily modified to apply 
to the spin matrices.

\section{Results}
\label{section-results}

\begin{theorem}
{\rm (Accardi in \cite{accardi})} 
Suppose  
$ Q $ and $ P $ are the standard self-adjoint 
realizations of the position and
momentum operators acting in the Hilbert 
space $ L^{2} (\mathbb{R}) $. 
Let $ S_{1}, S_{2} $ be bounded Borel subsets 
of $ \mathbb{R} $. 	
Then $ E_{Q} (S_{1}) E_{P} (S_{2}) $ is a trace class
operator, where $ E_{Q} $ (resp., $ E_{P} $) is the
projection valued measure on $ \mathbb{R} $ 
associated by the spectral theorem with 
the self-adjoint operator $ Q $ (resp., $ P $). 
Moreover,
$$
    Tr \big(  E_{Q} (S_{1}) E_{P} (S_{2}) \big) =
    \mu_{L} (S_{1}) \mu_{L} (S_{2}),
$$
where $ Tr $ denotes the trace of a trace class 
operator and $ \mu_{L} $ is a rescaling of Lebesgue 
measure. 
\end{theorem} 

We use this theorem to motivate a definition 
in the context of this paper. 

\begin{definition}
Let $A,B \in su(2) $ lie on 
the unit sphere, i.e.,
$ \mathrm{tr} \, A^{2} = \mathrm{tr} \, B^{2} = 1 $, 
with $ E_{A}, E_{B} $ being their projection valued 
measures. 
Then we say that $ A,B $ are {\rm Accardi complementary} 
if for all Borel subsets $ S_{1}, S_{2} $
of $ \mathbb{R} $ we have 
that 
\begin{equation}
\label{define-comp}
    \big\langle 
     E_{A} (S_{1} ) , E_{B} (S_{2}) 
    \big\rangle 
    =
    \mathrm{tr} \, 
    \big( E_{A} (S_{1} ) E_{B} (S_{2}) \big)
     = \mu_{B} (S_{1}) \mu_{B} (S_{2}),
\end{equation}
where $ \mu_{B} $ is the symmetric Bernoulli 
probability measure on $ \mathbf{R} $ supported on 
the  set  
$ \mathrm{Spec} \, A =  \mathrm{Spec} \, B = 
\{ -1, +1 \}$, that is
$ \mu_{B} ( \{ -1 \} ) = \mu_{B} ( \{ +1 \} ) = 1/2 $.
\end{definition}
This definition follows the 
original 
motivation of this concept
in \cite{accardi} where it is shown that
\eqref{define-comp} is equivalent to saying that 
measuring a value of $ A $ gives no further 
information on what the value of $ B $ will be
on a subsequent measurement of it, and {\em vice versa}.

We now want to find the projection valued measure 
for any $A \in su(2) $ with
norm $ 1 $. 
To do this we first use the isomorphism $ \Sigma  $
to write $ A = \alpha \cdot \sigma $ for a 
unique $ \alpha \in S^{2} $ in order 
to find its eigenvectors. 
\begin{prop}
Suppose that $ \alpha \in S^{2} $. 
Then a normalized eigenvector 
of the matrix $ \alpha \cdot \sigma $ 
with eigenvalue $ +1 $ is the column vector
\begin{equation}
\label{e-vector-for-plus}
\psi_{\alpha}^{+} :=
\dfrac{1}{2^{1/2} (1 + \alpha_{3} )^{1/2}}
\left( \begin{array}{l} 
            1 + \alpha_{3} 
            \\
            \alpha_{1} + i \alpha_{2} 
       \end{array} 
\right)
\in \mathbb{C}^{2} 
\quad \mathrm{if~} \alpha_{3} \ne -1.
\end{equation}
And a normalized eigenvector 
of the matrix $ \alpha \cdot \sigma $ 
with eigenvalue $ - 1 $ is the column vector 
\begin{equation}
\label{e-vector-for-minus}
\psi_{\alpha}^{-} :=
\dfrac{1}{2^{1/2} (1 - \alpha_{3} )^{1/2}}
\left( \begin{array}{l} 
- 1 + \alpha_{3} 
\\
\alpha_{1} + i \alpha_{2} 
\end{array} 
\right)
\in \mathbb{C}^{2} 
\quad \mathrm{if~} \alpha_{3} \ne +1.
\end{equation}
\end{prop}
\textbf{Proof:}
The eigenvalue equations (with unknowns $ x,y $) are  
\begin{equation}
\label{e-value-eqn}
\left( 
\begin{array}{cc} 
\alpha_{3} & \alpha_{1} - i \alpha_{2}  \
\\ 
\alpha_{1} + i \alpha_{2}  & -\alpha_{3} 
\end{array} 
\right) 
\left( \begin{array}{l} 
x  
\\
y
\end{array} 
\right)
= \pm 
\left( \begin{array}{l} 
x  
\\
y
\end{array} 
\right),  
\end{equation}
where the sign $ + $ (resp., $ - $) corresponds 
to eigenvalue $ +1 $ (resp., $ -1 $). 
Using the hypothesis $ || \alpha ||^{2} =1 $, one 
easily checks that the vectors in 
\eqref{e-vector-for-plus} and \eqref{e-vector-for-minus} 
satisfy \eqref{e-value-eqn} with the appropriate sign
and that they have norm $ 1 $. 
$ \quad \blacksquare $

\vskip 0.2cm 
Now we find the projection 
valued measure $ E_{\alpha \cdot \sigma} $
of $ \alpha \cdot \sigma $ for 
$ \alpha \in S^{2} $. 

\begin{theorem}
Suppose that $ \alpha \in S^{2} $. 
Then for the two non-trivial subsets of
$ \mathrm{Spec} \, ( \alpha \cdot \sigma ) = 
\{ -1 , +1 \} $ 
we have that 
\begin{equation}
\label{plus-projection}
E_{\alpha \cdot \sigma} (\{ +1 \}) = \dfrac{1}{2}
\left( 
\begin{array}{cc} 
1 + \alpha_{3} & \alpha_{1} - i \alpha_{2}  \
\\ 
\alpha_{1} + i \alpha_{2}  &  1 -\alpha_{3} 
\end{array} 
\right) 
=
\dfrac{1}{2} ( I + \alpha \cdot \sigma)
\end{equation}
and
\begin{equation}
\label{minus-projection}
E_{\alpha \cdot \sigma} (\{ -1 \}) = \dfrac{1}{2}
\left( 
\begin{array}{cc} 
1 - \alpha_{3} & - \alpha_{1} + i \alpha_{2}  \
\\ 
-\alpha_{1} - i \alpha_{2}  &  1 +\alpha_{3} 
\end{array} 
\right) 
=
\dfrac{1}{2} ( I - \alpha \cdot \sigma)
\end{equation}
\end{theorem}
\textbf{Proof:}
Using Dirac notation, we have 
$ E_{\alpha \cdot \sigma} (\{ +1 \}) 
= | \psi_{\alpha}^{+} \rangle 
\langle \psi_{\alpha}^{+} |$, 
where $ \psi_{\alpha}^{+} $ is given in 
\eqref{e-vector-for-plus}.
Then one can 
calculate the 
matrix in the middle of \eqref{plus-projection} 
directly from this formula by multiplying the 
column matrix $ | \psi_{\alpha}^{+} \rangle $ 
by the row matrix 
$\langle \psi_{\alpha}^{+} | = 
| \psi_{\alpha}^{+} \rangle^{*} $. 
Or one can verify that the expression on the 
rightmost side of \eqref{plus-projection} is a
projection whose range is spanned by
$  \psi_{\alpha}^{+} $. 

But the most elegant proof is to note that
$ E_{\alpha \cdot \sigma} (\{ +1 \})  
= \chi_{1}  (\alpha \cdot \sigma) $
by spectral theory, where
$ \chi_{1} $ is the characteristic function 
of the set $\{ 1 \} $, and then to use 
interpolation with Lagrange polynomials 
which says every function $ f $ of a $ 2 \times 2 $ 
matrix $ A $ is equal to a polynomial $ p $
of degree at most $ 1 $
of $ A $, where $ p $ must satisfy 
$ p = f $ on the spectrum of $ A $. 
Therefore we have $ \chi_{1} (x) = ax + b $ for all 
$ x \in \mathrm{Spec} \, ( \alpha \cdot \sigma ) = 
\{ -1, +1 \} $ for 
unknown coefficients $ a,b $. 
This gives the equations
$$
   1 = \chi_{1} (1) = a + b \quad \mathrm{and}
   \quad 0 = \chi_{1} (-1) = a - b,  
$$
whose solution clearly is $ a = b = 1/2 $. 
Thus, $ \chi_{1} (x) = (1/2) ( 1 + x )  = : p_{1}(x)$ 
for  all 
$ x \in \{ -1, +1 \} $. 
Finally,
$ 
E_{\alpha \cdot \sigma} (\{ +1 \})  
= \chi_{1}  (\alpha \cdot \sigma) 
= p_{1} (\alpha \cdot \sigma) = 
\frac{1}{2} ( I + \alpha \cdot \sigma)
$.

The expressions in \eqref{minus-projection} can be 
proved similarly or, even quicker, 
by using that 
$ E_{\alpha \cdot \sigma} (\{ -1 \}) = 
I - E_{\alpha \cdot \sigma} (\{ +1 \}) $. 
$ \quad \blacksquare $

\vskip 0.2cm
Of course, 
$ E_{\alpha \cdot \sigma} (\{-1,  +1 \}) = I$
and 
$ E_{\alpha \cdot \sigma} (\{ \emptyset \}) = 0$ by 
spectral theory, where 
$ \emptyset $ denotes the empty set 
and $ 0 $ denotes the zero operator.  

Also, 
the expressions in \eqref{plus-projection} and 
\eqref{minus-projection} indicate 
that the singularities
in \eqref{e-vector-for-plus} and 
\eqref{e-vector-for-minus} are removable. 

\begin{prop}
\label{prop-3.2}
For each subset $ S \subset \{ -1, +1\} $, we
have 
$ \mathrm{tr} \, E_{\alpha \cdot \sigma} ( S ) =
\mu_{B} (S)  $ for all $ \alpha \in S^{2} $.
In short, 
$ \mathrm{tr} \circ E_{\alpha \cdot \sigma}  =
\mu_{B}  $ for all $ \alpha \in S^{2} $. 
\end{prop}
\textbf{Proof:}
There are four such subsets $ S $. 
So we prove this in each of those four cases. 
For $ S = \emptyset $ the result is trivial.
For $ S= \{ -1, +1\} $ we have 
$$
\mathrm{tr} \, E_{\alpha \cdot \sigma} ( S ) = 
\mathrm{tr} \, E_{\alpha \cdot \sigma} ( \{ -1, +1 \} )=
\mathrm{tr} \, I = 1 = \mu_{B} ( \{ -1, +1 \} )
= \mu_{B} ( S ). 
$$

Finally, we obtain immediately from \eqref{plus-projection} 
and \eqref{minus-projection} that 
$ \mathrm{tr} \, E_{\alpha \cdot \sigma} (\{ +1 \}) 
=
\mathrm{tr} \, E_{\alpha \cdot \sigma} (\{ - 1 \}) 
= 1/2 $ for all $ \alpha \in S^{2} $. 
Since $ \mu_{B} (\{ +1 \}) = \mu_{B} (\{ -1 \}) = 1/2 $, 
we have proved the remaining two cases as well. 
$ \quad \blacksquare $

\vskip 0.2cm 
This proposition shows how spectral theory 
and the state $\mathrm{tr}$ give 
rise to the probability measure $ \mu_{B} $. 
We now are ready for the main result of this paper.
\begin{theorem}
\label{main-result}
	Suppose $ \alpha , \beta \in S^{2} $. 	
	Then $ \alpha \cdot \sigma, 
	\, \beta \cdot \sigma  $ 
	is Accardi complementary if and only if 
	$ \langle \alpha \cdot \sigma, 
	\beta \cdot \sigma \rangle = 
	\langle \alpha , \beta \rangle = 0 $,
	where the first inner product is that of $ su(2) $ 
	and the second is the standard inner product 
	on $ \mathbb{R}^{3} $. 
\end{theorem}
\textbf{Proof:}
It suffices 
to compute 
$ \mathrm{tr} \big(E_{\alpha \cdot \sigma} (S_{1}) 
E_{\beta \cdot \sigma} (S_{2}) \big)$ for all subsets
$ S_{1}, S_{2} $ of $ \{ -1, +1\} $.
If $ S_{1} = \emptyset $, then 
$$
\mathrm{tr} \big( E_{\alpha \cdot \sigma} (S_{1}) 
E_{\beta \cdot \sigma} (S_{2})  \big) = 
\mathrm{tr} ( 0 \, 
E_{\beta \cdot \sigma} (S_{2})  ) =
\mathrm{tr} ( 0 ) = 0
$$
while 
$$
\mu_{B} (S_{1}) \mu_{B} (S_{2}) = 0 \cdot 
\mu_{B} (S_{2})
= 0. 
$$
This shows that \eqref{define-comp} 
holds for this case.
The case $ S_{2} = \emptyset $ is proved similarly. 
If $ S_{1} = \{ -1, +1 \} $, then  
$$
\mathrm{tr} \big( E_{\alpha \cdot \sigma} (S_{1}) 
E_{\beta \cdot \sigma} (S_{2})  \big) = 
\mathrm{tr} ( I \, 
E_{\beta \cdot \sigma} (S_{2})  ) =
\mathrm{tr} ( E_{\beta \cdot \sigma} (S_{2})  ) 
$$
while 
$$
\mu_{B} (S_{1}) \mu_{B} (S_{2}) = 1 \cdot 
\mu_{B} (S_{2}) =  \mu_{B} (S_{2}). 
$$
So \eqref{define-comp} holds in this case by 
Proposition~\ref{prop-3.2}. 
The case $ S_{2} = \{ -1, +1 \} $ is proved similarly. 

We now consider the cases when both $ S_{1} $ 
and $ S_{2} $ contain exactly one element. 
In all of these remaining cases we have that 
$
\mu_{B} ( S_{1} ) \, \mu_{B} ( S_{2} ) = (1/2) (1/2) =
1 / 4. 
$

For the case $ S_{1} = S_{2} = \{ +1 \} $, we have 
\begin{align*}
 &\mathrm{tr} \big(E_{\alpha \cdot \sigma} (S_{1}) 
E_{\beta \cdot \sigma} (S_{2}) \big)
= 
\mathrm{tr} \big(E_{\alpha \cdot \sigma} (\{ +1 \}) 
E_{\beta \cdot \sigma} (\{ +1 \}) \big)
\\
&= \dfrac{1}{4} \, \mathrm{tr} 
\big( 
( I + \alpha \cdot \sigma) ( I + \beta \cdot \sigma) \big)
=
\dfrac{1}{4}  \, \mathrm{tr} 
\big(
I + \alpha \cdot \sigma + \beta \cdot \sigma 
+ (\alpha \cdot \sigma) (\beta \cdot \sigma) 
\big)
\\
&=
\dfrac{1}{4} (1 + 0 + 0 + \langle \alpha, \beta \rangle)
=
\dfrac{1}{4} ( 1 + \langle \alpha, \beta \rangle).  
\end{align*}
Here in the fourth equality we used 
\eqref{traceless} and \eqref{unitary}. 
Therefore, \eqref{define-comp} holds in this case if and 
only if $ \langle \alpha, \beta \rangle =0 $. 

For the case $ S_{1} = S_{2} = \{ -1 \} $, we have 
\begin{align*}
&\mathrm{tr} \big(E_{\alpha \cdot \sigma} (S_{1}) 
E_{\beta \cdot \sigma} (S_{2}) \big)
= 
\mathrm{tr} \big(E_{\alpha \cdot \sigma} (\{ -1 \}) 
E_{\beta \cdot \sigma} (\{ -1 \}) \big)
\\
&= \dfrac{1}{4} \, \mathrm{tr} 
\big( 
( I - \alpha \cdot \sigma) ( I - \beta \cdot \sigma) \big)
=
\dfrac{1}{4}  \, \mathrm{tr} 
\big(
I - \alpha \cdot \sigma - \beta \cdot \sigma 
+ (\alpha \cdot \sigma) (\beta \cdot \sigma) 
\big)
\\
&=
\dfrac{1}{4} ( 1 + \langle \alpha, \beta \rangle).  
\end{align*}
So \eqref{define-comp} holds in this case if and 
only if $ \langle \alpha, \beta \rangle =0 $. 

Next we consider the case $ S_{1} = \{ +1 \} $ 
and $ S_{2} = \{ -1 \} $. 
Then we have 
\begin{align*}
&\mathrm{tr} \big(E_{\alpha \cdot \sigma} (S_{1}) 
E_{\beta \cdot \sigma} (S_{2}) \big)
= 
\mathrm{tr} \big(E_{\alpha \cdot \sigma} (\{ +1 \}) 
E_{\beta \cdot \sigma} (\{ -1 \}) \big)
\\
&= \dfrac{1}{4} \, \mathrm{tr} 
\big( 
( I + \alpha \cdot \sigma) ( I - \beta \cdot \sigma) \big)
=
\dfrac{1}{4}  \, \mathrm{tr} 
\big(
I + \alpha \cdot \sigma - \beta \cdot \sigma 
- (\alpha \cdot \sigma) (\beta \cdot \sigma) 
\big)
\\
&=
\dfrac{1}{4} ( 1 - \langle \alpha, \beta \rangle).  
\end{align*}
Again \eqref{define-comp} holds in this case if and 
only if $ \langle \alpha, \beta \rangle =0 $. 

The remaining case $ S_{1} = \{ -1 \} $ 
and $ S_{2} = \{ +1 \} $ is proved similarly 
to the previous case. 
$ \quad \blacksquare $

\begin{definition}
	A subset 
	of the unit sphere 
	in $ su(2) $ is 
	{\rm Accardi complementary} if every 
	subset of it with exactly $ 2 $ elements is Accardi complementary. 
\end{definition}

\begin{corollary}
Let $ \alpha, \beta, \gamma $ be an orthonormal
basis of $ \mathrm{R}^{3} $. 
Then any pair of matrices in the set
$\{  \alpha \cdot \sigma, \beta \cdot \sigma, 
\gamma \cdot \sigma \}$ is 
Accardi complementary. 
In particular the triple of Pauli matrices 
$\{ \sigma_{1}, \sigma_{2} , \sigma_{3} \} $ is 
Accardi complementary. 
\end{corollary}
This result encodes the common folk knowledge in quantum 
physics that says measuring the spin of a spin $ 1/2 $
particle in some direction 
gives no information about subsequent spin
measurements in any orthogonal direction.
See \cite{accardi} for more on this point. 

\begin{corollary}
	Every subset of the unit sphere 
	in $ su(2) $ with $ 4 $ or more 
	elements is not Accardi complementary. 
\end{corollary}

\section{Concluding Remarks}
\label{conclusions}

There are properties of sets in mathematics which 
are of {\em finite type}, that is, a set has 
the property if and only if every finite subset 
of it has the property. 
The property of linear independence for subsets
of a vector space is a property of finite type.

There are other properties of sets 
which are of {\em unary type},
that is, a set has the property if and only if 
every subset with exactly $ 1 $ element has the property.
The property of a set of vectors 
in a normed vector being normalized 
is a property of unary type.

There are other properties of sets which are
of {\em binary type}, that is, 
a set has the property if and only if 
every subset with exactly 
$ 2 $ elements has the property.
The property that a set of vectors in a Hilbert space
is orthogonal is a property of binary type. 
We have shown that Accardi complementarity is a 
meaningful property of binary type by giving examples of
triples each of whose pairs is 
Accardi complementary.  

Next, let us note that the main theorem  
of this paper indicates that the {\em symmetric} 
Bernoulli probability measure $ \mu_{B} $ 
is in some sense a natural measure
on $ \mathrm{Spec} \, (\alpha \cdot \sigma) $ 
for any $ \alpha \in S^{2} $. 
Of course, $ \mu_{B} $ is the unique probability 
measure on 
$ \mathrm{Spec} \, (\alpha \cdot \sigma) = \{-1, +1\}$ 
with maximum entropy or, equivalently, with 
minimal information. 
This is its relevant property for this topic. 
It is a curious fact that $ \mu_{B} $ is also the 
normalized Haar measure on the finite 
multiplicative group $ \{-1, +1\} $. 

Theorem~\ref{main-result} clearly should generalize 
to any irreducible representation of $ su(2) $, 
that is to say in terms of quantum physics,
to any particle with spin $ n \hbar /2 $ with
$ n \ge 0 $ being an integer. 

Finally, after the preliminary version of 
this paper was finished, I learned 
about the results in \cite{accardi-lu}, where 
a stronger notion of complementarity is introduced 
for arbitrary sets of observables 
and examples of these are given and studied. 

\vskip 0.2cm 
\centerline{\bf Acknowledgment}
\vskip 0.2cm 
I thank Luigi Accardi for bringing 
the papers \cite{accardi} and \cite{accardi-lu} 
to my attention.

\end{document}